\documentstyle[prl,aps,epsfig,preprint]{revtex}

\def\norc #1 {\ {\Vert} #1 {\Vert_{c}}}
\def\norN #1 {\ {\Vert} #1 {\Vert_{N}}}
\def\nor #1 {\ {\Vert} #1 {\Vert}}

\def \Ninf#1 {  {\Vert {#1} \Vert_\infty } }
\def \NLip#1 {  {\Vert {#1} \Vert_\theta } }
\def \Norm#1 {  {\Vert {#1} \Vert } }

\begin{document}
\draft
\title{Control of Decoherence: Dynamical Decoupling versus Quantum
Zeno Effect
\\  - a case study for trapped ions - }

\author{S. Tasaki$^{[{\rm a}]}$, A. Tokuse$^{[{\rm a}]}$,
P. Facchi$^{[{\rm b}]}$ and S. Pascazio$^{[{\rm b}]}$ }
\address{$^{[a]}$Department of Applied Physics and
Advanced  Institute for
Complex Systems,
\\ Waseda University, 3-4-1 Okubo,
Shinjuku-ku, Tokyo 169-8555, JAPAN
\\
$^{[b]}$Dipartimento di Fisica, Universit\`a di
Bari, \\
and Istituto Nazionale di Fisica Nucleare, Sezione di Bari \\
I-70126 Bari, ITALY}

\date{\today}
\maketitle

\begin{abstract}

{\baselineskip=17pt

The control of thermal decoherence via dynamical decoupling and
via the quantum Zeno effect (Zeno control) is investigated for a
model of trapped ion, where the dynamics of two low lying
hyperfine states undergoes decoherence due to the thermal
interaction with an excited state. Dynamical decoupling is a
procedure that consists in periodically driving the excited state,
while the Zeno control consists in frequently measuring it. When
the control frequency is high enough, decoherence is shown to be
suppressed. Otherwise, both controls may accelerate decoherence. }

\end{abstract}



\newpage

\narrowtext

\section{INTRODUCTION}

The theory of quantum information and computation has provided
various promising ideas such as substantially faster algorithms
than their classical counterparts and very secure
cryptography\cite{Review}. Examples are Shor's factorizing
algorithm\cite{Shor} and Grover's search algorithm\cite{Grover},
where several computational states are simultaneously described by
a single wave function and parallel information processing is
carried out by unitary operations. Moreover, some of the basic
steps have already been experimentally realized: Basic operations
for quantum computation were realized with trapped
ions\cite{Wineland,Hughes} and with the nuclear spins of organic
molecules\cite{Cory}. Shor's algorithm for factorizing $Z=15$ was
investigated with the nuclear spins of organic
molecules\cite{NMRshor}.

The essential ingredient for the efficiency of quantum algorithms
and cryptography is the principle of superposition of states. As
pointed out e.g., by Unruh\cite{Unruh}, the loss of purity (i.e.,
decoherence) of states would deteriorate the performance,
particularly in the case of large scale computations or of
long-distance communications. Thus, the information carried by a
quantum system has to be protected from decoherence. So far,
several schemes have been proposed, such as the use of quantum
error-correcting codes\cite{ErrorCorrecting}, the use of
decoherence free subspaces and/or noiseless
subsystems\cite{NoiselessSub} and the quantum dynamical
decoupling\cite{Lloyd1,Lloyd2,Lloyd3,Vitali,Lidar}.

The quantum dynamical decoupling was proposed by Viola and
Lloyd\cite{Lloyd1}, where the system is periodically driven with
period $T_c$ in an appropriate manner so that the target subsystem
is decoupled from the environment. It was
shown\cite{Lloyd1,Lloyd2} that a complete decoupling is achieved
in the $T_c \to 0$ limit, or the limit of infinitely fast control.
The procedure is simpler than the other methods because one only
has to  periodically drive the system. However, as it is not
possible to achieve the $T_c \to 0$ limit, its performance for
nonvanishing $T_c$ should be investigated. Such studies were
carried out for a two-level system in an environment via a
system-energy-preserving interaction\cite{Lloyd1} and for a
harmonic oscillator coupled with an environment\cite{Vitali}.
Here, we will provide one more example, namely a model of a
trapped ion used in Ref.\cite{Wineland}. This model explicitly
involves a unitary operation for the quantum-state manipulation,
which was not included in the previous models.

The key ingredient of dynamical decoupling is the continuous
disturbance of the system, which suppresses the system-environment
interaction. As already pointed out by Viola and
Lloyd\cite{Lloyd1}, the situation is similar to the so-called
quantum Zeno effect, where frequent measurements of a system
suppress quantum transitions\cite{QZEReview,WinelandZeno,QZEbxl}
(for recent reviews, see \cite{17bis}). This phenomenon is more
general than originally thought: a nontrivial time evolution may
occur in the case of frequent measurements under an appropriate
setting. Namely, when the measurement process is described by a
multidimensional projection operator, frequent measurements
restrict the evolution within each subspace specified by the
projection operator and a superselection rule dynamically
arises\cite{FPsuper}. Therefore, if one can design the measurement
process so that different superselection sectors (defined by the
given measurements) are coupled by the interaction between a
target system and the environment, the system-environment
interaction can be suppressed by frequent measurements. We refer
to such a decoherence control as a {\it quantum Zeno control}.
Since, in case of the quantum Zeno experiment by Itano et
al.\cite{WinelandZeno,QZEbxl}, the measurement process was
realized as a dynamical process, namely the optical pulse
irradiation, it is interesting to compare the two procedures (the
quantum Zeno control and the quantum dynamical decoupling) for a
model of trapped ion. This is one of the objectives of this
article.

This article will be organized as follows. In Sec. \ref{sec.general},
the quantum dynamical decoupling and the quantum Zeno control are
briefly reviewed. In Sec. \ref{sec.model}, we introduce a model of the
trapped ion, which takes into account the unitary Rabi oscillation and
thermal decoherence. The dynamical decoupling and Zeno controls of this
model are discussed, respectively, in Secs. \ref{sec.bang} and
\ref{sec.Zeno}. After discussing the cases of infinitely fast controls,
the effects of the finiteness of the control period are investigated.
It is shown that both controls may accelerate decoherence if they are
not sufficiently fast. This implies the necessity of a careful design
of the control and a careful study of the timescales involved. The last
section is devoted to the summary and discussion.

\section{Quantum Dynamical Decoupling and Quantum Zeno Control}
\label{sec.general}

\subsection{System}

The total system consists of a target system and a reservoir and
its Hilbert space ${\cal H}_{\rm tot}$ is the tensor product of
the system Hilbert space, ${\cal H}_S$, and the reservoir Hilbert
space, ${\cal H}_B$: ${\cal H}_{\rm tot}={\cal H}_S \otimes {\cal
H}_B$. The total Hamiltonian $H_{\rm tot}$ is the sum of the
system part $H_S \otimes {\bf 1}_B$, the reservoir part ${\bf 1}_S
\otimes H_B$ and their interaction $H_{SB}$, which is
responsible for decoherence:
\begin{equation}
H_{\rm tot} =H_S \otimes {\bf 1}_B + {\bf 1}_S
\otimes H_B+ H_{SB}(t) \ . \label{Hamiltonian1}
\end{equation}
The operators ${\bf 1}_S$ and ${\bf 1}_B$ are
the identity operators, respectively, in the Hilbert spaces
${\cal H}_S$ and ${\cal H}_B$, and the operators
$H_S$ and $H_B$ act,
respectively, on
${\cal H}_S$ and ${\cal H}_B$.
Here, in order to discuss controls in an interaction picture,
a time-dependent interaction is considered.

Since, in general, the reservoir state is mixed, it is convenient
to describe the time evolution in terms of density matrices. In
the case of a quantum state manipulation, the initial state
$\rho(0)$ is set to be a tensor product of a system initial state
$\sigma(0)$ and a reservoir (usually equilibrium) state $\rho_B$:
$\rho(0)=\sigma(0)\otimes \rho_B$. The system state $\sigma(t)$ at
time $t$ is given by the partial trace of the state $\rho(t)$ of
the whole system with respect to the reservoir degrees of freedom:
$\sigma(t) \equiv {\rm tr}_B \rho(t)$. When $\sigma(t)$ is not
unitarily equivalent to $\sigma(0)$ for a given class of initial
states, decoherence is said to appear. The purpose of the control
is to suppress such decoherence. For the decoherence
control, it is sufficient to consider only those initial states
which are relevant to the quantum state manipulation in question,
but not all states.

\subsection{Quantum Dynamical Decoupling}

Here we slightly generalize the arguments of Ref.\cite{Lloyd2} (see
also
\cite{Vitali}). This control is carried out via a time dependent {\it
system} Hamiltonian $H_c(t)$:
\begin{equation}
H(t) =H_{\rm tot} + H_c(t) \otimes {\bf 1}_B \ , \label{bb-Hamiltonian}
\end{equation}
where $H_c(t)$ is designed so that $U_c(t) \equiv {\cal T}
\exp\left\{ -i \int_0^t H_c(s) ds\right\}$ satisfies

\noindent (A) \ \  $U_c(t)$ is periodic with period $T_c$; $U_c(t+T_c)
=U_c(t)$.

\noindent (B) \ \  $\int_0^{T_c} dt \bigl( U_c^\dag(t) \otimes {\bf 1}
_B\bigr) H_{SB}(t+s)
\bigl( U_c(t) \otimes {\bf 1}_B \bigr) ={\rm O}(T_c^{1+\epsilon}) \ ,
\quad (0<\epsilon
\le1, T_c: \ {\rm small}, \forall s)$.

Going to the interaction picture where $H_c(t)$ is unperturbed,
the density matrix at time $T=N T_c$ with an initial state
$\rho(0)$ is given by $\rho(T) = U_{\rm tot}(NT_c) \rho(0) U_{\rm
tot}^\dag (NT_c)$ where
\begin{equation}
U_{\rm tot}(NT_c) = {\cal T} \exp\left\{ -i \int_0^{NT_c}
{\widetilde H}_{\rm tot}(s) ds\right\} =
\prod_{m=1}^N \left[ {\cal T}
\exp\left\{ -i \int_{(m-1)T_c}^{m T_c} {\widetilde H}_{\rm tot}(s)
ds\right\}\right]
\end{equation}
and ${\widetilde H}_{\rm tot}(t)= \bigl(U_c^\dag(t) \otimes
{\bf 1}_B\bigr)
 H_{\rm tot} \bigl(U_c(t) \otimes {\bf 1}_B\bigr)$.
A standard Magnus expansion of the time
ordered exponential\cite{Magnus} leads to
\begin{equation}
 {\cal T} \exp\left\{
-i \int_{(m-1)T_c}^{m T_c} {\widetilde H}_{\rm tot}(s) ds\right\}
= e^{-i[{\bar H}^{(0)}_m+{\bar H}^{(1)}_m+\cdots]T_c}
\end{equation}
where ${\bar H}^{(0)}_m\equiv {1\over T_c}
\int_{(m-1)T_c}^{m T_c} {\tilde H}_{\rm tot}(s) ds$ and the rest terms
${\bar H}^{(j)}_m$ are of order of $T_c^j$ ($j=1,2,\cdots$).
By assumption (B), one has ${\bar H}^{(0)}_m={\bar H}^{(0)}+{\rm
O}(T_c^\epsilon)$ where
\begin{equation}
{\bar H}^{(0)} = {\bar H}_S \otimes {\bf 1}_B + {\bf 1}_S
\otimes H_B
\ ,
\end{equation}
${\bar H}_S \equiv {1\over T_c}
\int_0^{T_c} dt U_c^\dag(t) H_{S} U_c(t)$ and they
are independent of $T_c$ because  $U_c(t)$ is $T_c$-periodic.
Therefore, in the limit $T_c\to 0$ while keeping
$T=N T_c$ constant, one obtains
\begin{equation}
U_{\rm tot}(T) \simeq
\left[ 1- i {\bar H}^{(0)} {T\over N} + {\rm O}\left({1\over
N^{1+\epsilon}}\right)
\right]^N \stackrel{N \to \infty}{\longrightarrow} e^{-i{\bar H}_S T}
\otimes e^{-i H_B T}
\ .
\end{equation}
In short, as a result of the infinitely fast control, the
system-reservoir coupling is eliminated and, thus,
decoherence is suppressed.
Note that if one designs $H_c$
so that
\begin{equation}
U_c(t)\equiv g_j \qquad \left( {j-1\over M}T_c\le t < {j\over M}T_c
\ ; j=1,\cdots,M\right) ,
\label{untaryControl}
\end{equation}
where $\{ g_j \}$ is a set of unitary operators acting on ${\cal H}_S$,
${\bar H}_S$ becomes
\begin{equation}
{\bar H}_S = {1\over M} \sum_{j=1}^M g_j^\dag H_S g_j
\label{UnitaryBangH}
\end{equation}
The relation between the dynamical decoupling with the
prescription (\ref{untaryControl}) and a symmetry group was
discussed in Ref.\cite{Lloyd3}.

\subsection{Quantum Zeno control}

Now we turn to the Zeno control by adapting the argument of
Ref.\cite{FPsuper}.
This control is performed by frequent measurements of
the system.
The most general measurement
is described by a projection operator acting on the
density matrix:
\begin{equation}
\rho \to {\hat P}\rho \equiv \sum_n \left(P_n\otimes {\bf 1}_B \right)
\rho \left(P_n\otimes {\bf 1}_B\right) \ .
\label{measurement}
\end{equation}
where $\{ P_n\}$ is a set of orthogonal Hermitian projection
operators acting on ${\cal H}_S$. In the following, we restrict
ourselves to the case where the measuring apparatus does not
``select'' different outcomes (nonselective
measurement)\cite{Schwinger} and the projection operators are
complete; $\sum_n P_n={\bf 1}_S$. As in the dynamical decoupling,
the measurement is designed so that

\noindent (C) \ \ ${\hat P}H_{SB}(t) = \sum_n \left(P_n\otimes {\bf 1}
_B\right)
H_{SB}(t) \left(P_n\otimes {\bf 1}_B\right) =0$.

The Zeno control consists in performing repeated nonselective
measurements at times $t=n T_c$ ($n=0,1,2,\cdots$).
Between successive measurements, the system evolves via $H_{\rm
tot}$. In terms of the Liouville operator ${\cal L}_{\rm tot}$
defined by ${\cal L}_{\rm tot}\rho \equiv [H_{\rm tot},\rho]=
H_{\rm tot}
\rho -
\rho H_{\rm tot}$, the density matrix $\rho(NT_c)$ after $N+1$
measurements with an initial state $\rho(0)$ is given by
\begin{equation}
\rho(NT_c) = \prod_{m=1}^N \left\{{\hat P} {\cal
T}e^{-i\int_{(m-1)T_c}^{mT_c} {\cal L}_{\rm tot}(t) dt} {\hat
P}\right\}
\rho(0)
\ .
\end{equation}
Assumption (C) yields
\begin{eqnarray}
{\hat P}\int_{(m-1)T_c}^{mT_c} {\cal L}_{\rm tot}(t)
dt {\hat P}\rho &=& {\hat P} \int_{(m-1)T_c}^{mT_c} [ {\hat P}(H_{\rm
tot}(t)),
\rho] dt \nonumber \\
&=& T_c {\hat P} [ {\bar H}'_S
\otimes {\bf 1}_B + {\bf 1}_S
\otimes H_B, \rho] \equiv T_c {\overline {\cal L}}^{(0)} \rho
\ , \nonumber
\end{eqnarray}
and, thus, in the limit $T_c\to 0$ while keeping $T=N T_c$
constant, we get
\begin{eqnarray}
\rho(NT_c) &\simeq& {\hat P} \prod_{m=1}^N \left\{\left(
{\bf 1}-i {\hat P} \int_{(m-1)T_c}^{mT_c} {\cal L}_{\rm tot}(t)
{\hat P}
dt\right) \right\}
\rho(0)
\simeq {\hat P}\left\{ \left(1- i{\overline {\cal
L}}^{(0)} {T\over N}\right) \right\}^N  \rho(0) \nonumber
\\
&\to& {\hat P} e^{-i{\overline {\cal
L}}^{(0)} T} \rho(0) = {\hat P} \left( e^{-i{\bar
H}'_S T} \otimes e^{-i H_B T} \rho(0) e^{i{\bar H}'_S T} \otimes e^{i
H_B T} \right)
\ ,
\end{eqnarray}
where the controlled system Hamiltonian ${\bar H}'_S$ is given by
\begin{equation}
{\bar H}'_S \equiv \sum_n P_n H_{S} P_n \ . \label{QZcontrolH}
\end{equation}
Hence, as a result of infinitely frequent measurements, the
system-reservoir coupling is eliminated and, thus, decoherence is
suppressed. Note the similarity between the controlled system
Hamiltonians for a particular dynamical decoupling,
(\ref{UnitaryBangH}), and for the Zeno control,
(\ref{QZcontrolH}). This is not a mere coincidence. Indeed, one
can show that, by enlarging the Hilbert space so that the original
measurement process is expressed by a dynamical process in the
larger space, the two controls are equivalent. This will be
discussed in detail elsewhere. However, throughout this article,
the dynamical decoupling refers to a situation where the
evolution is coherent (unitary) and the Zeno control to a
situation where the evolution involves incoherent processes such
as measurements.

\section{A Two-level System with Thermal Decoherence}
\label{sec.model}

\subsection{Model}

\begin{figure}
\begin{center}
\epsfig{file=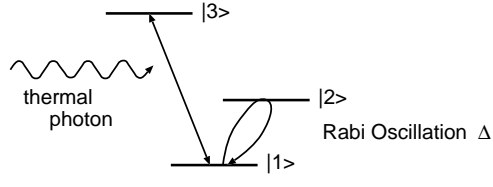,width=6.5cm}
\end{center}
\caption{A schematic picture of the system.
$|1\rangle$, $|2\rangle$ and $|3\rangle$ are the lower hyperfine,
upper hyperfine and excited states, respectively.}
\label{fig:fig1}
\end{figure}

We consider the model of a trapped Be ion used in
Ref.\cite{Wineland} (see also \cite{WinelandZeno}). Here we assume
that the ion is at rest and consider only the dynamics of the
hyperfine states. Then, the main mechanism of decoherence can be
attributed to the emission and absorption of thermal photons
associated  with transitions to nearby excited states. For the
sake of simplicity, one of the hyperfine states is assumed to
couple electromagnetically with a nearby excited state (see Fig.\
\ref{fig:fig1}),
the polarization of the photon is neglected and only the rotating
terms (i.e., the slowly varying terms in the interaction picture)
are taken into account (rotating wave approximation)
for the driven parts.

Let $|1\rangle$, $|2\rangle$ and $|3\rangle$ be the lower
hyperfine, upper hyperfine and excited states, respectively, and
$a_{\bf k}$ the annihilation operator of a photon with wave vector
${\bf k}$ and energy $\omega_k = c|{\bf k} |$ ($c$: speed of
light). Then, the Hamiltonian is given by
\begin{eqnarray}
H_{\rm tot}(t) &=&H_S(t) \otimes {\bf 1}_B  + {\bf 1}_S \otimes H_B+
\lambda H_{SB}
\ , \label{ModelHamiltonian} \\
H_S(t) &=& \omega_2 |2\rangle \langle 2| + \omega_3 |3\rangle
\langle 3|
+ \lambda^2 \{\Delta e^{i\omega_2' t} |1\rangle \langle 2| + (h.c.)\}
\ , \label{Hs}\\
H_B &=& \int d^3 {\bf k} \ \omega_k \ a_{\bf k}^\dag
a_{\bf k} \ , \label{HB} \\
H_{SB} &=& \int d^3 {\bf k} V_k  \ \left(|1\rangle \langle 3|
+|3\rangle \langle 1|\right)
\otimes \left( a_{\bf k}^\dag + a_{\bf k}\right)
\ . \label{Hin}
\end{eqnarray}
The bare energies $\omega_2$ and $\omega_3$ of states $|2\rangle$
and $|3\rangle$ are measured from that of the lower hyperfine
state $|1\rangle$ ($\omega_1=0$). The third term of (\ref{Hs})
represents the rf-control of the Rabi oscillation and the
amplitude $\lambda^2
\Delta$ corresponds to the Rabi frequency. Because of the Lamb
shift, the energy difference between the two hyperfine states
$\omega_2'$ is different from $\omega_2$ and the frequency of the
irradiated field should be so tuned that it resonates with
$\omega_2'$. The function $V_k$ is assumed to behave like
\begin{equation}
V_k = \sqrt{v_0^2 c^3 \over {8\pi^2 \omega_k}}e^{-{\omega_k \over
2\omega_c}} \ , \label{formFac}
\end{equation}
with a cut-off frequency $\omega_c$ and a dimensionless strength
$v_0$. The dimensionless coupling constant $\lambda$ measures the
relative order of magnitude of each term and is of order
$\sqrt{\gamma_e/\omega_3}$, where $\gamma_e$ is the inverse
lifetime of the excited state. For the system in \cite{Wineland},
typical order of magnitudes of the frequencies are
$\gamma_e/\omega_3 \simeq 10^{-8}$ and (Rabi frequency)$/\omega_3
\simeq 10^{-10}$, which imply that $\lambda \simeq 10^{-4}$ and
the Rabi frequency is of order $\lambda^2$. Also it is reported
that the decoherence time is longer than the Rabi period by two
orders of magnitude \cite{Wineland}.

It is convenient to move to a rotating frame with the aid of the
unitary operator
$$
U_R(t) = \exp\left\{i\left(\lambda^2 \delta |1\rangle \langle 1|
+ \omega_2 |2\rangle \langle 2| +\omega_3
|3\rangle \langle 3|\right)t \right\}\otimes
\exp\left\{i \omega_3' t \int d^3{\bf k} \ a_{\bf k}^\dag
a_{\bf k}   \right\} ,
$$
where $\lambda^2 \delta= \omega_2-\omega_2'$ and $\omega_3'=\omega_3-
\lambda^2\delta$.
Then the transformed Hamiltonian $H_{\rm tot}^R$ is
\begin{equation}
H_{\rm tot}^R \equiv i {\partial U_R(t) \over \partial t}
U_R^\dag(t) + U_R(t) H_{\rm tot}(t)
U_R^\dag(t) = \lambda^2 H_S^R \otimes {\bf 1}_B  + {\bf 1}_S \otimes
H_B^R + \lambda H_{SB}^R(t)
\ , \label{rModelHamiltonian}
\end{equation}
\begin{eqnarray}
H_S^R &=& - \delta |1\rangle \langle 1|
+ \{\Delta |1\rangle \langle 2| + (h.c.)\}
\ , \label{rHs}\\
H_B^R &=& \int d^3 {\bf k} \ (\omega_k-\omega_3') \
a_{\bf k}^\dag
a_{\bf k} \ , \label{rHB} \\
H_{SB}^R(t) &=& \int d^3 {\bf k} V_k  \ |1\rangle \langle 3|
\otimes \left(a_{\bf k}^\dag + e^{-2i\omega_3' t} a_{\bf k}\right)
+ (h.c.)
\end{eqnarray}
This is our starting point.

\subsection{Decoherence}\label{sec.decoherence}

We consider the time evolution starting from an initial state
given by the tensor product of a system initial state and the
reservoir equilibrium state with inverse temperature $\beta$:
\begin{eqnarray}
i{\partial \rho(t) \over \partial t} &=& \left\{{\cal L}_B +
\lambda {\cal L}_{SB}(t)
+ \lambda^2 {\cal L}_S\right\} \rho(t) \label{LNmodel} \\
\rho(0) &=& \sigma(0)\otimes \rho_B \\
\rho_B &=& {1\over Z} \exp(-\beta H_B) \label{BathEq}
\end{eqnarray}
where $Z$ is the normalization constant and the operators ${\cal
L}_B$, ${\cal L}_{SB}$ and ${\cal L}_S$ are defined by
\begin{eqnarray}
{\cal L}_B \rho &\equiv& [{\bf 1}_S\otimes H_B^R, \rho] \ , \qquad
{\cal L}_{SB}(t) \rho \equiv [H_{SB}^R(t), \rho] \ , \qquad {\cal L}_S
\rho \equiv [ H_S^R \otimes {\bf 1}_B, \rho] \ .
\end{eqnarray}
Since the time scale of the  quantum state manipulation is of the
same order of magnitude as the Rabi period ($\sim \lambda^{-2}$)
and is very long compared with $1/\omega_3$ ($\sim \lambda^{0}$),
the process is well described by the van Hove limit
approximation\cite{vanHove,vanHoveLeb}. The starting point is the
decomposition of the Liouville equation (\ref{LNmodel}) with the
aid of a projection operator
\begin{eqnarray}
{\cal P}\rho \equiv \left({\rm tr}_B \rho\right) \otimes \rho_B
\end{eqnarray}
where ${\rm tr}_B$ stands for the partial trace over the reservoir
degrees of freedom and $\rho_B$ is the equilibrium reservoir state
(\ref{BathEq}).
Then, in the limit $\lambda \to 0$ while keeping
$\tau =\lambda^2 t$ constant, the reduced density matrix
$\sigma \equiv {\rm tr}_B \rho$ is found to
satisfy\cite{vanHove,vanHoveLeb,Yuasa}
\begin{eqnarray}
{\partial \sigma \over \partial \tau} = -i\Lambda \sigma \ ,
\label{decohOp}
\end{eqnarray}
where we have used ${\cal P}{\cal L}_B {\cal P}=0$ and
\begin{eqnarray}
-i\Lambda \sigma &=& -i[ H_S^R, \sigma]- \int_0^{+\infty} dt \
\lim_{\lambda \to 0}{\rm
tr}_B
\left\{ {\cal L}_{SB}\left({\tau\over \lambda^2}\right) e^{-i{\cal
L}_B t} {\cal L}_{SB}\left({\tau\over \lambda^2}-t\right) e^{i{\cal
L}_B t} \
\sigma (\tau)\otimes \rho_B \right\}
\nonumber \\
&=& -i\left[\delta' |3\rangle \langle
3|+\Delta \bigl( |1\rangle \langle 2|+(h.c)\bigr),\sigma \right]
\nonumber \\ &&+ \gamma_e |1\rangle \langle 3|\sigma|3\rangle \langle 1|
+ \gamma_d |3\rangle \langle 1|\sigma|1\rangle \langle 3|
- \left\{ {\gamma_d\over 2}|1\rangle \langle 1|
+ {\gamma_e\over 2} |3\rangle \langle 3|,\sigma
\right\} \ , \label{decohEvo}
\end{eqnarray}
with $\{ \ , \}$ the anti-commutator.
In the above,
the limit $\lambda \to 0$ should be understood to drop terms
which oscillate with frequencies $\sim \lambda^{-2}$
(cf. \cite{vanHoveLeb}).
The parameter $\delta$ is chosen as
\begin{eqnarray}
\delta = p.v. \int_{-\infty}^{+\infty} d\omega {\kappa_d(\omega)\over
\omega-\omega_3'}
\end{eqnarray}
and $\delta'$, $\gamma_d$ and $\gamma_e$ are given by
\begin{eqnarray}
\delta' &=& - p.v. \int_{-\infty}^{+\infty}  d\omega
{\kappa_e(\omega)\over
\omega-\omega_3'}
\\
\gamma_d &=& 2\pi \kappa_d(\omega_3'),\qquad \gamma_e=2\pi
\kappa_e(\omega_3')\ ,
\end{eqnarray}
where
\begin{equation}
\kappa_d(\omega)=\kappa_e(\omega) e^{-\beta \omega}=\int d^3 {\bf k}
V_k^2 {\delta(\omega_k-\omega)+ e^{\beta\omega_k}
\delta(\omega_k+\omega)
\over e^{\beta
\omega_k}-1}=\frac{v_0^2 }{2\pi}{ \omega\; e^{-|\omega|/\omega_c}
\over e^{\beta \omega}-1}
\label{eq:formfactors}
\end{equation}
are the thermal spectral density functions (form factors). The symbol
$p.v.$ in front of the integrals indicates Cauchy's principal value.

In terms of the matrix elements $\sigma_{ij} \equiv
\langle i|\sigma|j \rangle$, one has
$\sigma_{21}={\overline \sigma_{12}}$ and
\begin{eqnarray}
{\partial \sigma_{11} \over \partial \tau}
&=& -i \Delta \{ \sigma_{21}-\sigma_{12}\} - \gamma_d
\sigma_{11}+\gamma_e \sigma_{33} \\
{\partial \sigma_{12} \over \partial \tau}
&=& -i \Delta \{ \sigma_{22}-\sigma_{11}\} - {\gamma_d
\over 2} \sigma_{12} \\
{\partial \sigma_{22} \over \partial \tau}
&=& i \Delta \{ \sigma_{21}-\sigma_{12}\} \\
{\partial \sigma_{33} \over \partial \tau}
&=& \gamma_d \sigma_{11}-\gamma_e \sigma_{33} \ .
\end{eqnarray}

\begin{figure}
\begin{center}
\epsfig{file=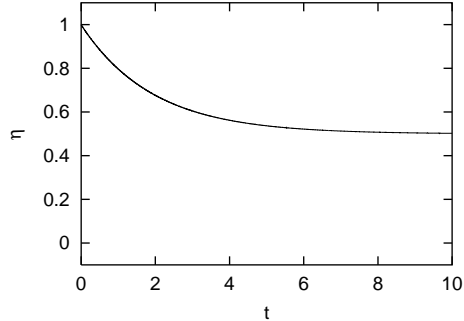,width=6.5cm}
\end{center}
\caption{Time evolution of the purity $\eta$ of the
target states. The time unit on the horizontal axis is the
decoherence time $\gamma_d^{-1}$.}
\label{fig:fig2}
\end{figure}

The purity of the target states is measured by $\eta \equiv
\sigma_{11}^2 + \sigma_{22}^2 + 2 |\sigma_{12}|^2$,
as $\eta=1$ for pure superpositions of $|1\rangle$ and
$|2\rangle$, and $\eta<1$ for states involving the irrelevant
state $|3\rangle$ or mixed states.
In Fig.\ \ref{fig:fig2}, the evolution of the quantity
$\eta(t)$ starting from
$\sigma(0)=|1\rangle \langle 1|$ is shown for $\Delta=100
\gamma_d, \gamma_e=1000 \gamma_d$.
As time goes on, the purity $\eta$ of the target states
is lost, or
decoherence takes place. One clearly sees that the decoherence
time scales as $1/\gamma_d$.

\section{Control of Thermal Decoherence via Dynamical Decoupling}
\label{sec.bang}

\begin{figure}
\begin{center}
\epsfig{file=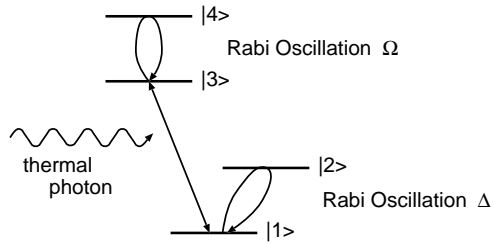,width=6.5cm}
\end{center}
\caption{A schematic picture of the system under the
quantum dynamical decoupling control.}
\label{fig:fig3}
\end{figure}

We consider a dynamical decoupling control of the thermal
decoherence discussed in the previous section. Since decoherence
arises from the transition between the states $|1\rangle$ and
$|3\rangle$ associated with absorption and emission of photons, it
is expected to be suppressed if $|3\rangle$ does not contribute to
the $|1\rangle$-$|2\rangle$ dynamics. So we consider a control via
the Rabi oscillation between the state $|3\rangle$ and a higher
excited state $|4\rangle$ (cf. Fig.\ \ref{fig:fig3}),
\noindent
which is described by
\begin{eqnarray}
H_c(t)&=& \Omega e^{i(\omega_4+\xi) t} |3\rangle \langle 4| + (h.c.)
\ ,
\label{eq:controlt}
\end{eqnarray}
where $\omega_4$ is the energy of the state $|4\rangle$, and
$\Omega$ and $\xi$ are real parameters. Since there exists one
more state, a term $\omega_4 |4\rangle \langle 4|\otimes {\bf
1}_B$  should be added to the Hamiltonian $H_S$. As a result, the
following Liouvillian should be added
\begin{eqnarray}
\left\{\Delta{\cal L}_S+{\cal L}_c(t)\right\} \rho \equiv
\left[\left(\omega_4 |4\rangle \langle 4|+H_c(t)\right) \otimes
{\bf 1}_B \ , \ \rho\right] \ ,
\end{eqnarray}
and the evolution equation reads
\begin{eqnarray}
{\partial \rho \over \partial t}&=& \left\{{\cal L}_B +
\Delta{\cal L}_S + {\cal L}_c(t) + \lambda {\cal L}_{SB}(t) +
\lambda^2 {\cal L}_S\right\} \rho(t) \ . \label{LNmodel+c}
\end{eqnarray}

\subsection{Ideal dynamical decoupling}

The evolution operator $U_c(t)$ generated by the control
Hamiltonian $H_c(t)$ is given by
\begin{eqnarray}
U_c(t)&=&{\cal T} \exp\left(-i\int_0^t dt' H_c(t')\right) \nonumber \\
&=& e^{-i(\omega_4+\xi)|4\rangle \langle 4| t}
\ \exp\left[-i\left\{-(\omega_4+\xi)|4\rangle \langle 4|
+\Omega (|3\rangle \langle 4|+(h.c))\right\} t  \right] \ .
\end{eqnarray}
When restricting to a subspace spanned by $|3\rangle$ and
$|4\rangle$, the second factor is a sum of two oscillating
projection operators with frequencies
\begin{eqnarray}
{\tilde \omega}_{\pm} = {1\over 2} \left(-\omega_4-\xi
\pm \sqrt{(\omega_4+\xi)^2+4\Omega^2}\right)
\end{eqnarray}
and the first factor is a sum of a time-independent and an
oscillating projection operators, the frequency of the latter
being $\omega_4+\xi$. Thus, $U_c(t)$ is a sum of four oscillating
terms with frequencies $\pm {\tilde \omega}_+, \pm {\tilde
\omega}_-$ and is $T_c$-periodic provided both ${\tilde \omega}_+
T_c$ and ${\tilde \omega}_- T_c$ are integer multiples of $2\pi$.

Under this prescription, $\bigl(U_c^\dag(t) \otimes {\bf 1}_B
\bigr) H_{SB} \bigl(U_c(t)\otimes {\bf 1}_B\bigr)$ is
a sum of terms proportional to $e^{\pm i {\tilde \omega}_+ t},
e^{\pm i {\tilde \omega}_- t}, e^{\pm 2i \omega_3' t}e^{\pm i
{\tilde \omega}_+ t}, e^{\pm 2i \omega_3' t} e^{\pm i {\tilde
\omega}_- t},$ and its average vanishes in the $T_c\to 0$ limit:
\begin{eqnarray}
\lim_{T_c\to 0}{1\over T_c}\int_0^{T_c} dt \bigl(U_c^\dag(t) \otimes
{\bf 1}_B
\bigr) H_{SB} \bigl(U_c(t)\otimes {\bf 1}_B\bigr) = 0 \ .
\end{eqnarray}
Therefore, the general argument of Sec.\ref{sec.general} shows
that the coupling  between the system and the reservoir is
suppressed in the limit $T_c\to 0$ (that is ${\tilde
\omega}_{\pm}\to\infty$) while keeping $T=NT_c$ constant.
Moreover, the system obeys the Hamiltonian
\begin{eqnarray}
{\bar H}_s &=& {1\over T_c}\int_0^{T_c} dt U_c^\dag(t)
\left\{ \omega_4 |4\rangle \langle 4| + \lambda^2 H_S^R \right\}
U_c(t) \nonumber \\
&=& \omega_4 \sum_{s=\pm}{\tilde \omega}_s^2{\left(\Omega |3\rangle +
{\tilde \omega}_s |4\rangle \right)\left(\Omega \langle 3| + {\tilde
\omega}_s \langle 4| \right) \over (\Omega^2 +{\tilde \omega}_s^2)^2 }
+\lambda^2 H_S^R \ . \label{eq:barHs}
\end{eqnarray}
Therefore, the target system spanned by $|1\rangle$ and
$|2\rangle$ is free from decoherence and performs ideal Rabi
oscillations.

It is interesting to see the relation between the dynamical
decoupling and  a dynamical quantum Zeno effect due to a
``continuous" measurement\cite{FPsuper}.
For the particular choice (\ref{eq:controlt}), it is possible to
eliminate the explicit time dependence of the control Hamiltonian
$H_c$, by going to another rotating frame with the aid of the
unitary operator
\begin{eqnarray}
U_R^b(t)= \exp\left\{i (\omega_4+\xi) |4\rangle \langle 4|t
\right\} \otimes {\bf 1}_B \ ,
\label{eq:URb}
\end{eqnarray}
and, then, the transformed density matrix ${\bar \rho}= U_R^b(t)
\rho U_R^{b\dag}(t)$ obeys
\begin{eqnarray}
{\partial {\bar \rho} \over \partial t}&=& \left\{{\cal L}_B
+\Delta{\cal L}_S'+ {\cal L}_c' + \lambda {\cal L}_{SB} +
\lambda^2 {\cal L}_S\right\} {\bar \rho} \label{LNmodel+cb}
\end{eqnarray}
where the transformed control Liouvillian is
\begin{eqnarray}
\left\{\Delta{\cal L}_S'+{\cal L}_c'\right\} \rho =
\left[\left(-\xi |4\rangle \langle 4|+ \Omega \{|3\rangle \langle
4|+ (h.c.) \}\right) \otimes {\bf 1}_B \ , \ \rho\right] \ .
\end{eqnarray}
In this picture, state $|4\rangle$ of energy $-\xi$ is coupled by
a constant coupling $\Omega$ to state $|3\rangle$. The
short-period limit $T_c\to 0$ corresponds to the strong coupling
limit $\Omega, \xi \to
\infty$, because ${\tilde\omega}_{\pm}T_c$ must be integer multiples of
$2\pi$. But this is just the case of a dynamical quantum Zeno effect
due to a continuous measurement\cite{FPsuper}, where $H_{\rm meas}=-\xi
|4\rangle \langle 4|+ \Omega \{|3\rangle \langle 4|+ (h.c.) \}$ plays
the role of a measurement Hamiltonian. One can show that, in the limit
of strong coupling, a dynamical superselection rule arises, the Hilbert
space is split into Zeno subspaces and the system Hamiltonian is given
again by (\ref{QZcontrolH}), where the projections $P_n$s, defining the
Zeno subspaces, are the eigenprojections of $H_{\rm meas}$
\cite{FPsuper}. This is a consequence of an interesting relation
between strong-coupling regime and adiabatic evolution \cite{Frasca}.

The eigenprojections of $H_{\rm meas}$ are given by
$P_\pm=|\pm\rangle\langle\pm|$, where
\begin{equation}
|\pm\rangle=\frac{\Omega |3\rangle + { \omega}_\pm
|4\rangle}{\sqrt{\Omega^2 +{\omega}_\pm^2}} \ ,
\label{eq:pmstates}
\end{equation}
with eigenvalues
\begin{eqnarray}
\omega_\pm = {1\over 2} \left( -\xi \pm \sqrt{\xi^2+4\Omega^2}
\right) \ .
\label{eq:omegapm}
\end{eqnarray}
Therefore from (\ref{QZcontrolH}) one gets
\begin{eqnarray}
{\bar H}_s &=& \sum_{s=\pm}P_s\left( -\xi |4\rangle \langle 4| +
\lambda^2 H_S^R \right)P_s= -\xi \sum_{s=\pm}{\omega}_s^2{\left(\Omega
|3\rangle + { \omega}_s |4\rangle \right)\left(\Omega \langle 3| +
{\omega}_s \langle 4| \right) \over (\Omega^2 +{\omega}_s^2)^2 }
+\lambda^2 H_S^R \ ,
\end{eqnarray}
which is nothing but the Hamiltonian (\ref{eq:barHs}) under the
transformation (\ref{eq:URb}). We therefore see that, in this
particular case, dynamical decoupling is completely equivalent to
the dynamical Zeno effect.

\subsection{Nonideal dynamical decoupling}
\label{NonidealBang}

Here we consider the case of nonvanishing $T_c$ and solve the
evolution equation (\ref{LNmodel+c}). As pointed out in
\cite{Lloyd1,Vitali}, the ideal dynamical decoupling
is achieved when the control frequency $2\pi/T_c$ is higher than
the threshold frequency $\omega_c$ in the system-reservoir
interaction $H_{SB}$ and, thus, we consider the case where
$2\pi/T_c = {\rm O}(\lambda^0)$. Then, the slow process which is
relevant to the quantum state manipulation is well described by
the van Hove limit approximation\cite{vanHove,vanHoveLeb,Yuasa}.

We consider the evolution in the rotated frame (\ref{LNmodel+cb}).
Note that, since the transformation $U_R^b$ in (\ref{eq:URb}) does
not affect the evolution of the states $|j \rangle$ ($j=1,2$) and
the field variable, one has $\langle i|{\rm tr}_B \rho |j\rangle =
\langle i|{\rm tr}_B {\bar \rho} |j\rangle$ ($i,j=1,2$). By the
standard procedure of the van Hove limit
approximation\cite{vanHove,vanHoveLeb}, in the limit $\lambda \to 0$
while keeping $\tau =\lambda^2 t$ constant, one obtains
\begin{eqnarray}
{\partial \sigma \over \partial \tau} = -i \Lambda_{\rm B} \sigma
\end{eqnarray}
where $\sigma(\tau)\equiv {\rm tr}_B
{\bar \rho}$ and
\begin{eqnarray}
 -i \Lambda_{\rm B} \sigma
&=&
-i\left[\sum_{s=\pm}\delta_s |s\rangle \langle
s|+\Delta \bigl( |1\rangle \langle 2|+(h.c)\bigr),\sigma \right]
\nonumber \\
&&+ \sum_{s=\pm} \left( \gamma_e^s |1\rangle \langle s|\sigma|s\rangle
\langle 1| + \gamma_d^s |s\rangle \langle 1|\sigma|1\rangle \langle s|
\right)
- \left\{ {\gamma_d^B\over 2}|1\rangle \langle 1|
+ \sum_{s=\pm}{\gamma_e^s\over 2} |s\rangle \langle s|,\sigma
\right\} \ .
\end{eqnarray}
The states $|\pm \rangle$ are the normalized linear combinations of the
states $|3\rangle$ and $|4\rangle$ given by (\ref{eq:pmstates}) and the
decay rates $\gamma_e^s$ and $\gamma_d^s$ ($s=\pm$) and $\gamma_d^B$
are given by
\begin{eqnarray}
\gamma_d^B=\gamma_d^+ + \gamma_d^-, \qquad \gamma_{d/e}^\pm = 2\pi
{|\omega_\mp|\over \omega_+ -\omega_-}
\kappa_{d/e}(\omega_3'+\omega_\pm)
, \label{eq:gammadpm}
\end{eqnarray}
where $\omega_\pm$ are the frequencies (\ref{eq:omegapm}) and
$\kappa_{d/e}(\omega)$ are the thermal form factors
(\ref{eq:formfactors}), extended to the whole real axis due to the
counter-rotating terms. (Incidentally, notice the exchange
symmetry $\kappa_e(\omega)=\kappa_d(-\omega)$ of the extended form
factors.)

The prefactors in the second equation in (\ref{eq:gammadpm}) are
nothing but the squares of the matrix elements between the
undressed state $|3\rangle$ and the dressed states $|\pm\rangle$
(\ref{eq:pmstates}):
\begin{equation}
{|\omega_\mp|\over \omega_+ -\omega_-}=\left|\langle 3
|\pm\rangle\right|^2.
\end{equation}
The explicit expressions of the Lamb shifts $\delta_s$ ($s=\pm$) of the
excited states are omitted since the relevant sector of the evolution
equation does not depend on them. Note that the parameter $\delta$ is
chosen so that the operator $\Lambda_{\rm B}$ does not contain a
commutator with $|1\rangle \langle 1|$.

\begin{figure}
\begin{center}
\epsfig{file=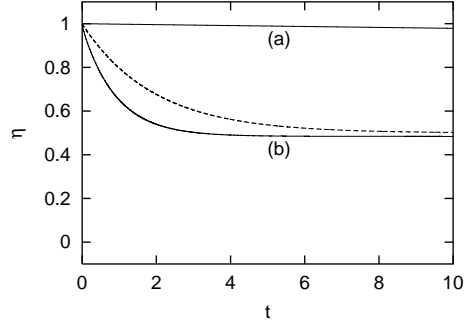,width=6.5cm}
\end{center}
\caption{\label{fig:fig4} Evolution of the purity $\eta$ of the
system state. The time unit in the horizontal axis is the
decoherence time $\gamma_d^{-1}$ for the uncontrolled case. (a)
Control frequency $|\omega_-|=150 \times \omega_3'$; (b) control
frequency $|\omega_-|=0.5 \times \omega_3'$. For comparison, the
behavior of $\eta$ without control is also displayed (broken
curve).}
\end{figure}

In terms of the matrix elements $\sigma_{ij} \equiv
\langle i|\sigma|j \rangle$, one has
\begin{eqnarray}
{\partial \sigma_{11} \over \partial \tau}
&=& -i \Delta \{ \sigma_{21}-\sigma_{12}\} - \gamma_d^B
\sigma_{11}+\sum_{s=\pm}\gamma_e^s \sigma_{ss} \\
{\partial \sigma_{12} \over \partial \tau}
&=& -i \Delta \{ \sigma_{22}-\sigma_{11}\} - {\gamma_d^B
\over 2} \sigma_{12} \\
{\partial \sigma_{22} \over \partial \tau}
&=& i \Delta \{ \sigma_{21}-\sigma_{12}\} \\
{\partial \sigma_{ss} \over \partial \tau}
&=& \gamma_d^s \sigma_{11}-\gamma_e^s \sigma_{ss} \ . \quad (s=\pm)
\end{eqnarray}
The evolution of the purity $\eta = \sigma_{11}^2+\sigma_{22}^2
+2|\sigma_{12}|^2$ of the target states is shown in Fig.\
\ref{fig:fig4} for different values of control parameters, where
the parameters in the control Hamiltonian are set to
$\xi=24\Omega/5$ (which gives $\omega_+=\Omega/5$ and $\omega_-=-5
\Omega$), the frequency-cutoff to $\omega_c=10 \omega_3'$, and the
other parameters are chosen so that one has $\Delta=100 \gamma_d$
and $\gamma_e=1000
\gamma_d$ for the uncontrolled case. As in the previous section,
the initial state is $\sigma(0)=|1\rangle \langle 1|$. Fig.\
\ref{fig:fig4} shows that
the dynamical decoupling control may accelerate decoherence if the
parameters are not appropriately tuned.

\begin{figure}
\begin{center}
\epsfig{file=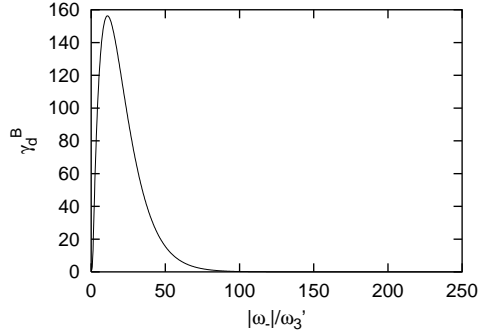,width=6.5cm}
\end{center}
\caption{Decoherence rate $\gamma_d^B$
vs control frequency $|\omega_-|/\omega_3'$.}
\label{fig:fig5}
\end{figure}

In Fig.\ \ref{fig:fig5}, the
control-frequency dependence of the decoherence rate $\gamma_d^B$
is shown. As can be seen in the figure, decoherence is first
enhanced, for small values of $|\omega_-| =
5\Omega$, and then suppressed for much larger
values of $|\omega_-|$. Since the decoherence rate $\gamma_d^+$
due to $|+\rangle$ is a monotonically decreasing function of
$|\omega_-|$, this can be understood as follows: In the
rotating frame where the
$|3\rangle$-$|4\rangle$ oscillation is eliminated, when $\Omega=0$,
state $|3\rangle$ is separated from the decay product state
$|1\rangle$ by an energy
$\omega_3'$. When $\Omega$ is turned on, state  $|3\rangle$ splits
into two dressed states $|\pm\rangle$ which are separated from
$|1\rangle$ by the energies
$\omega_3'+\omega_+ \equiv \omega_3'+|\omega_-|/25$ and
$\omega_3'-|\omega_-|$, respectively. The latter
state is closer to state
$|1\rangle$ than in the uncontrolled case and leads to a shorter
decoherence time provided $\omega_3'-|\omega_-| >0$. This is the
deterioration observed in Fig.\ \ref{fig:fig4}, case (b). On the
other hand, if $|\omega_-|$ exceeds a threshold energy $\omega_{th}
\equiv
\omega_3'$, the energy of state $|-\rangle$ becomes lower than
that of state $|1\rangle$.
In such a case, the counter-rotating term (which are now ``rotating")
{\em does} contribute to the decoherence rate.
Notice that now, being $\omega_3'-|\omega_-|<0$,
$\gamma_e^-$ is {\em smaller} than $\gamma_d^-$, as it should. Even
after $|\omega_-|$ has exceeded the threshold $\omega_{th}$,
$\gamma_d^-$ still increases with $|\omega_-|$, since the state
$|1\rangle$ is now unstable. And, finally, when the two dressed
states are sufficiently far apart from level $|1\rangle$, the decay
rates (and therefore decoherence) are suppressed because of the high
energy cut-off of the form factor (\ref{formFac}). Such values of
$|\omega_-| \sim 80
\omega_3'$ is extremely higher than the threshold
$\omega_{th}=\omega_3'$ (Fig.\ref{fig:fig5}) and involve
extremely short timescales \cite{FPintense}.

\section{Quantum Zeno Control of Thermal Decoherence}
\label{sec.Zeno}

\begin{figure}
\begin{center}
\epsfig{file=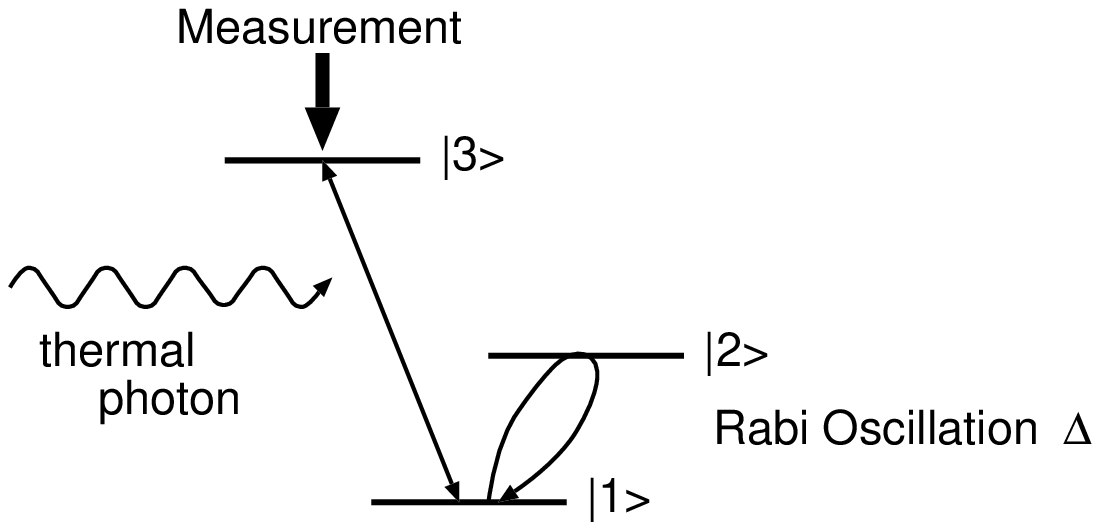,width=6.5cm}
\end{center}
\caption{A schematic picture of the system under the
quantum Zeno control.}
\label{fig:fig6}
\end{figure}

For the same reason as in the dynamical decoupling, we
disturb the evolution of $|3\rangle$ by repeated measurements
(Fig.\
\ref{fig:fig6}). The nonselective measurement of $|3\rangle$
causes the following change of the density matrix:
\begin{eqnarray}
\rho \to {\hat P}\rho &\equiv& \pi_3 \rho \pi_3
+ ({\bf 1}-\pi_3) \rho ({\bf 1}-\pi_3)
\end{eqnarray}
where $\pi_3$ is a projection operator acting on the
whole Hilbert space $\pi_3 \equiv |3\rangle \langle 3|
\otimes {\bf 1}_B$ and ${\bf 1}={\bf 1}_S\otimes
{\bf 1}_B$.
Then, the density matrix under the Zeno control is given by
\begin{equation}
\rho(NT_c) = \prod_{m=1}^N \left\{{\hat P} {\cal
T}e^{-i\int_{(m-1)T_c}^{mT_c} {\cal L}_{\rm tot}(t) dt} {\hat
P}\right\}
\rho(0)
\ ,
\end{equation}
where $T_c$ stands for the time interval between successive
measurements.

\subsection{Ideal Zeno control}

First we consider the case where $T_c\to 0$ while keeping $T=N
T_c$ constant. Then, as discussed in Sec.\ref{sec.general} and
because ${\hat P}H_{SB}=\pi_3 H_{SB}\pi_3 +({\bf 1}-\pi_3) H_{SB}
({\bf 1}-\pi_3)=0$, the state at time $T$ is given by
\begin{eqnarray}
\rho(T)
&=& {\hat P} \left( e^{-i{\bar H}'_S T} \otimes e^{-i H_B T} \rho(0)
e^{i{\bar H}'_S T} \otimes e^{i H_B T} \right)
\ ,
\end{eqnarray}
where the controlled system Hamiltonian ${\bar H}'_S$ is given by
\begin{eqnarray}
{\bar H}'_S &=&\lambda^2 \left\{ P_3 H_S^R P_3 +({\bf
1}_S-P_3)  H_S^R
({\bf 1}_S-P_3)\right\} \nonumber \\
&=&\lambda^2 \left\{  - \delta |1\rangle \langle 1|
+ \left(\Delta |1\rangle \langle 2| + (h.c.)\right) \right\}
\ , \label{QZcontrolH2}
\end{eqnarray}
with $P_3=|3\rangle \langle 3 |$. Hence, as a result of infinitely
frequent measurements of state $|3\rangle$, the system-reservoir
coupling is eliminated and, thus, decoherence is suppressed.

\subsection{Nonideal Zeno control}
\label{sec.Zenob}

As in the dynamical decoupling, we consider the case where
$T_c \sim 1/\omega_3' \sim \lambda^0$, so that the time evolution
is well described by the van Hove limit where $\lambda \to 0$
while keeping $\tau=\lambda^2 NT_c$ and $T_c$ constant. We are
here looking at the subtle effects on the decay rate arising from
the presence of a short-time quadratic (Zeno) region. Therefore,
it is important to notice that the standard
method\cite{vanHove,vanHoveLeb} is not applicable to the present
situation and the limit is evaluated as follows:

\noindent (1) \ Second order perturbation, up to $\lambda^2$, and
${\hat P}H_{SB}=0$  lead to
\begin{eqnarray}
{\hat P} {\cal
T}e^{-i\int_{mT_c}^{(m+1)T_c} {\cal L}_{\rm tot}(t) dt} {\hat
P}
&\simeq& {\hat P} e^{-i{\cal L}_B T_c} \biggl\{ {\bf 1} -
i\lambda^2 {\cal L}_S T_c \nonumber \\
&&-\lambda^2
\int_{mT_c}^{(m+1) T_c} dt \int_{mT_c}^t ds e^{i{\cal L}_B t}
{\cal L}_{SB}(t) e^{-i{\cal L}_B (t-s)} {\cal L}_{SB}(s)
e^{-i{\cal L}_B s} {\hat P}\biggr\} \ .
\end{eqnarray}

\noindent (2) \ In terms of the operator ${\cal K}_m$, defined as a
solution
of the operator equation
\begin{eqnarray}
{\hat P}\int_{mT_c}^{(m+1) T_c} dt \int_{mT_c}^t ds e^{i{\cal
L}_B t} {\cal L}_{SB}(t) e^{-i{\cal L}_B (t-s)} {\cal L}_{SB}(s)
e^{-i{\cal L}_B s}{\hat P} = {\hat P} \int_{mT_c}^{(m+1)T_c} dt
e^{i{\cal L}_B t} {\cal K}_m
e^{-i{\cal L}_B t}
\ ,
\end{eqnarray}

one has
\begin{eqnarray}
{\hat P} {\cal
T}e^{-i\int_{mT_c}^{(m+1)T_c} {\cal L}_{\rm tot}(t) dt} {\hat
P}
\simeq
{\hat P} {\cal
T}e^{-i\int_{mT_c}^{(m+1)T_c}  ({\cal L}_B +\lambda^2{\cal L}_S
-i\lambda^2 {\cal K}(t) ) dt}
+ {\rm O}(\lambda^3)
\ , \label{ZenoApp}
\end{eqnarray}

where ${\cal K}(t)={\cal K}_m$ for $m T_c \le t < (m+1)T_c$.

\noindent (3) \ With the aid of (\ref{ZenoApp})
and
${\hat p}\sigma \equiv P_3 \sigma P_3 + ({\bf 1}_S-P_3)\sigma ({\bf
1}_S-P_3)$, the final
reduced state

$\sigma(\tau)$ is given by
\begin{eqnarray}
\sigma(\tau) &=& \lim_{\lambda\to 0 \atop \tau=\lambda^2 NT_c
: {\rm finite}} {\rm tr}_B \left(
\prod_{m=1}^N \left\{{\hat P} {\cal
T}e^{-i\int_{(m-1)T_c}^{mT_c} {\cal L}_{\rm tot}(t) dt} {\hat
P}\right\}
\sigma(0)\otimes \rho_B
\right) \nonumber \\
&=& {\hat p} \lim_{\lambda\to 0 \atop \tau=\lambda^2
NT_c : {\rm finite}} {\rm tr}_B \rho^*(NT_c)
\ .
\end{eqnarray}

where $\rho^*(NT_c) = {\hat P} {\cal
T}e^{-i\int_{0}^{NT_c}  ({\cal L}_B +\lambda^2{\cal L}_S
-i\lambda^2 {\cal K}(t) ) dt} \sigma(0)\otimes \rho_B$.

\noindent (4) \ As $\rho^*(t)$ is a solution of
\begin{eqnarray}
{\partial \rho^*(t) \over \partial t} = -i({\cal L}_B +\lambda^2{\cal L}
_S
-i\lambda^2 {\cal K}(t) ) \rho^*(t) \ , \qquad \rho^*(0)=
\sigma(0)\otimes \rho_B
\ ,
\end{eqnarray}

the standard van Hove limit arguments\cite{vanHove,vanHoveLeb} show
that
$\sigma(\tau) ={\hat p}\sigma^*(\tau)$ and $\sigma^*$ satisfies
\begin{eqnarray}
{\partial \sigma^*(\tau) \over \partial \tau} = -i \ {\rm tr}_B \left\{
\left({\cal L}_S -i {\cal K}\right) \sigma^*(\tau) \otimes \rho_B
\right\}
\ , \qquad \sigma^*(0)=
\sigma(0) \ , \label{sig*Eq}
\end{eqnarray}

where the time-dependence of ${\cal K}$ is lost as a result of
the partial trace.

As in the previous sections, the parameter $\delta$ is chosen so
that the $|1\rangle \langle 1|$-term does not appear in the
evolution operator of $\sigma^*$.  In terms of the matrix elements
$\sigma_{ij}^* \equiv
\langle i|\sigma^*|j \rangle$, (\ref{sig*Eq}) reads
\begin{eqnarray}
{\partial \sigma_{11}^* \over \partial \tau}
&=& -i \Delta \{ \sigma_{21}^*-\sigma_{12}^*\} - \gamma_d^Z
\sigma_{11}^* +\gamma_e^Z \sigma_{33}^* \\
{\partial \sigma_{12}^* \over \partial \tau}
&=& -i \Delta \{ \sigma_{22}^* -\sigma_{11}^* \} - {\gamma_d^Z
\over 2} \sigma_{12}^* \\
{\partial \sigma_{22}^* \over \partial \tau}
&=& i \Delta \{ \sigma_{21}^* -\sigma_{12}^* \} \\
{\partial \sigma_{33}^* \over \partial \tau}
&=& \gamma_d^Z \sigma_{11}^* -\gamma_e^Z \sigma_{33}^* \ ,
\end{eqnarray}
where the decoherence rate $\gamma_d^Z$ and the inverse lifetime
$\gamma_e^Z$ of $|3\rangle$ are given by
\begin{eqnarray}
\gamma_d^Z &=& T_c \int_{-\infty}^\infty d \omega \; \kappa_d(\omega)\;
 {\rm sinc}^2 \left(\frac{\omega-\omega_3'}{2}
T_c\right) \label{gammadz}  \\
\gamma_e^Z &=& T_c \int_{-\infty}^\infty d \omega \; \kappa_e(\omega)\; {\rm sinc}^2
\left(\frac{\omega-\omega_3'}{2} T_c\right) \label{gammaez} \ ,
\end{eqnarray}
where $\kappa_{d/e}(\omega)$ are again the (extended) thermal form
factors ({\ref{eq:formfactors}) and
${\rm sinc}(x)=(\sin x) / x$. The decay rate $\gamma_d^Z$ in
({\ref{gammadz}) should be compared to $\gamma_d^B$
in (\ref{eq:gammadpm}). They express the (inverse) quantum Zeno
effect, given by pulsed or continuous measurement, respectively
\cite{17bis}.

Since the projection operator $\hat p$ does not affect the
$|1\rangle$-$|2\rangle$ sector, one has $\sigma_{ij}^*(\tau) =
\langle i|\sigma(\tau)|j \rangle$ for a class of initial states
where only the matrix elements $\langle i|\sigma(0)|j \rangle$
($i,j=1,2$) are nonvanishing.
Hence, $\eta = \sigma_{11}^{* 2} +
\sigma_{22}^{* 2} + 2|\sigma_{12}^*|^2 $ measures the purity of
the target states. Its evolution is shown in Fig.\ \ref{fig:fig7}
for different values of $2\pi/T_c$, where $\omega_c=10 \omega_3'$
and the other parameters are chosen so that one has $\Delta=100
\gamma_d$ and $\gamma_e=1000 \gamma_d$ for the uncontrolled case.
As in the previous sections, the initial state is
$\sigma(0)=|1\rangle
\langle 1|$. Fig.\ \ref{fig:fig7} shows that the Zeno
control may accelerate decoherence if the parameters are not
appropriately chosen. This can be seen more clearly in the
control-frequency dependence of the decoherence rate $\gamma_d^Z$,
which is shown in Fig.\ \ref{fig:fig8}. When the control frequency
$2\pi/T_c$ belongs to a certain range, decoherence is enhanced.

The enhancement of decoherence is qualitatively similar to the
case of the dynamical decoupling. However, the high frequency
behavior of the decoherence rate and its peak values are quite
different. The high-frequency decoherence rates $\gamma_d^B$ and
$\gamma_d^Z$, respectively, for the dynamical decoupling and Zeno
control, are approximated by
\begin{eqnarray}
\gamma_d^B \simeq {\omega_+ \gamma_e \omega_c\over
\omega_3'(\omega_+-\omega_-)} {|\omega_-|\over
\omega_c} e^{-{|\omega_-|\over \omega_c}} \ , \qquad
\gamma_d^Z \simeq {\gamma_e \over \left({2\pi\over \omega_3'
T_c}\right)}
\left({\omega_c\over \omega_3'}\right)^2 \ .
\end{eqnarray}
Therefore, $\gamma_d^B$ decays exponentially for large $|\omega_-|$
because of the exponential cut-off of the form factor and may take
a maximum of order $\omega_+ \gamma_e \omega_c/
\{e (\omega_+-\omega_-)\omega_3'\}\sim 140$.
On the other hand, $\gamma_d^Z$ decays polynomially for large
$2\pi/T_c$  and $\gamma_d^Z$ could be much larger than $\gamma_d^B$
because $\gamma_e \omega_c^2/ \omega_3'^2 \sim 10^5$ is
very large.
\begin{figure}
\begin{center}
\epsfig{file=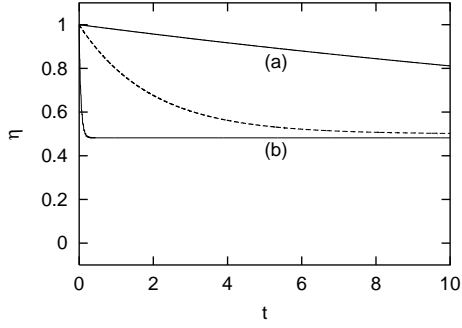,width=6.5cm}
\end{center}
\caption{Evolution of the purity $\eta$ of the target states.
The time unit on the horizontal axis is the decoherence time
$\gamma_d^{-1}$ for the uncontrolled case. (a) Control frequency
$2\pi/T_c=5\times 10^6 \times \omega_3'$; (b) control frequency
$2\pi/T_c= 0.5 \times
\omega_3'$. For comparison, the behavior of $\eta$ without control
is also displayed by a broken curve.}
\label{fig:fig7}
\end{figure}
\begin{figure}
\begin{center}
\epsfig{file=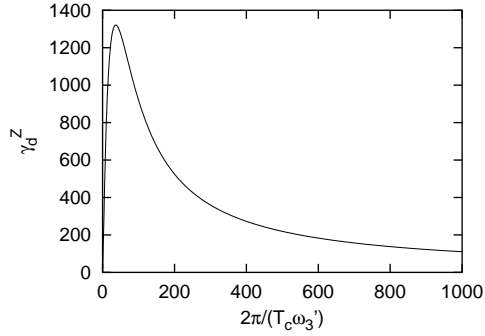,width=6.5cm}
\end{center}
\caption{Decoherence rate $\gamma_d^Z$ vs control
frequency $2\pi/(T_c \omega_3')$.}
\label{fig:fig8}
\end{figure}

\section{CONCLUSIONS}
\label{sec.conclusions}

In this article, we have studied the dynamical decoupling and Zeno
controls for a model of trapped ions, where decoherence appears in
the dynamics of the hyperfine states due to emission and
absorption of thermal photons associated with the transition
between the lower hyperfine and an excited state. By very rapidly
driving or very frequently measuring the excited state,
decoherence is shown to be suppressed. However, if the frequency
of the controls are not high enough, the controls may accelerate
the decoherence process and may deteriorate the performance of the
quantum state manipulation.

The acceleration of decoherence is analogous to the inverse Zeno
effect, namely the  acceleration of the decay of an unstable state
due to frequent measurements\cite{antiZeno}. In the original
discussion of the Zeno
effect\cite{QZEReview,WinelandZeno,QZEbxl,17bis}, very frequently
repeated measurements of an unstable state is shown to slow down
its decay. But, if the  duration between two successive
measurements is not short enough, the frequent measurements may
accelerate the decay. This is the inverse Zeno effect. Obviously,
this situation precisely corresponds to the increase of
decoherence observed in this article. Moreover, since a very
intense field is used for the dynamical decoupling
control, the decrease of the decoherence time is also a
consequence of the decrease of the lifetime of the unstable states
due to the intense field\cite{FPintense}.

There is room for improvement and further analysis: a number of
neglected effects can be considered, such as the role of
counter-rotating terms and Fano states, the influence of the other
atomic states, the primary importance of the relevant timescales,
and so on. These aspects will be discussed elsewhere.

\acknowledgments

{\small The authors are grateful to Professors I. Antoniou, B.
Misra, A. Takeuchi, I. Ohba, H. Nakazato, L. Accardi, T. Hida, M.
Ohya, K. Yuasa, and N. Watanabe for fruitful discussions and
comments. This  work is supported by a Grant-in-Aid for Scientific
Research (C) from JSPS and by a Grant-in-Aid for Scientific
Research of Priority Areas ``Control of Molecules in Intense Laser
Fields'' from the Ministry of Education, Culture, Sports, Science
and Technology of Japan. }

\end{document}